\documentclass[%
 reprint,
 amsmath,amssymb,
 aps,
]{revtex4-2}

\usepackage[english]{babel}

\usepackage{tikz}
\usetikzlibrary{angles,quotes}
\usetikzlibrary{calc}
\usetikzlibrary{patterns}

\usepackage{xcolor}
\usepackage{verbatim}

\usepackage{graphicx}
\usepackage{dcolumn}
\usepackage{bm}

\newcommand{\jfac}{\mathcal{J}}
\newcommand{\sigv}{\sigma_A v_{\mathrm{rel}}}
\newcommand{\vrel}{v_\mathrm{rel}}

\begin{document}

\preprint{APS/123-QED}

\title{Optimal observing strategies  for velocity-suppressed dark matter annihilation}

\author{Nolan Smyth}
\email{nwsmyth@ucsc.edu}
\author{Gabriela Huckabee}%
 \email{ghuckabe@ucsc.edu}
 \author{Stefano Profumo}%
 \email{profumo@ucsc.edu}
\affiliation{
    Department of Physics, University of California, Santa Cruz, 
    1156 High Street, Santa Cruz, CA 95064, USA
}
\affiliation{
    Santa Cruz Institute for Particle Physics, 
    1156 High Street, Santa Cruz, CA 95064, USA
}

\date{\today}

\begin{abstract}
Numerous particle models for the cosmological dark matter feature a pair-annihilation rate that scales with powers of the relative velocity between the annihilating particles. As a result, the annihilation rate in the central regions of a dark matter halo can be significantly lower than at the halo's  periphery for particular ambient gravitational potentials. While this might be offset by an increasing dark matter pair number density in the inner halo, it raises the question: what angular region for dark matter models with velocity-suppressed annihilation rates optimizes signal-to-noise? Here, we consider simplified background models for galactic and extragalactic targets and demonstrate that the optimal observing strategy varies greatly case-by-case. Generally, a bright central source warrants an annular region of interest, while a flatter background warrants as large as possible an angular region, possibly including the central regions.
\end{abstract}

\maketitle


\section{Introduction}
\label{sec:introduction}

The production of Standard Model particles from the annihilation of dark matter (DM) particles provides a detectable signal with which to indirectly probe DM models (for reviews, see e.g. \cite{particledatagroupReviewParticlePhysics2020, profumoDarkMatterIndirect2020, slatyerTASILecturesIndirect2017}). Dwarf spheroidal galaxies (dSphs) are of particular interest in indirect detection searches as they are generally DM-dominated and have a low astrophysical foreground \cite{strigariDarkMatterDwarf2018}. Other prime targets are the Galactic center region and nearby galaxies such as M31 \cite{doCosmicrayTransportGammaray2020, mcdanielMultiWavelengthAnalysisAnnihilating2018} and M87 \cite{profumoCosmicRayDarkMatter2011,profumoDarkMatterIndirect2020, slatyerTASILecturesIndirect2017}.

The total photon flux from a specific astrophysical object due to DM annihilation is proportional to that object's \textit{J-factor}, also sometimes called the \textit{astrophysical factor}. The J-factor is determined by the astrophysical properties of the object, such as its DM density profile. In calculating the predicted photon flux from DM annihilation, often one assumes a velocity-independent cross section, which, in turn, implies a velocity-independent J-factor. However, in cases where the velocity-independent $s$-wave channel is subdominant to the velocity-dependent $p$-wave and $d$-wave channels, the J-factor must effectively include the velocity-dependent contribution from the cross section. As shown by \citep{boardVelocitydependentJfactorsAnnihilation2021}, in the general case of a velocity-dependent cross section, the J-factor scales as the moments of the DM velocity distribution. They show that even in this general case, the J-factor is strongly correlated with the DM density and weakly correlated with the velocity dispersion. 

However, we show in this work that for simple cases of $p$-wave and $d$-wave dominated annihilation, the velocity-dependent J-factor implies that {\em the optimal signal-to-noise ratio is generally achieved for a different observation strategy than in the velocity-independent, $s$-wave dominated case}. For example, we find an annular field of view to be superior to a disk when the background gamma-ray signal is sharply peaked at the center of a DM halo. This is due to two factors. First, the velocity distribution of DM particles is peaked off-center in the halo, resulting in a boost to the velocity-dependent J-factor in this region. Second, the benefits of including the high-density region at the center of the halo in the J-factor are outweighed by the detriment of a large background.

In this study we consider two broad classes of observational targets: (1) a generic {\em extragalactic target} and (2) the Milky Way {\em Galactic Center}. The latter has been the subject of much debate due to the extended gamma-ray excess within the inner region of the bulge. Annihilating dark matter may be responsible for this excess and is the focus of this paper, but unresolved gamma-ray pulsars and other faint baryonic sources have also been put forth as viable explanations \cite{hooperDarkMatterAnnihilation2011, daylanCharacterizationGammaRaySignal2016, collaborationFermiLATObservationsHighEnergy2016, abazajianAstrophysicalDarkMatter2014}. The extragalactic class, which includes nearby galaxies such as M31, M87, and M33, is of interest because the central bulge and stellar disk are resolvable as two distinct components, something that is not possible in the Milky Way Galactic Center due to bright disk contamination. We include in this category local dwarf spheroidal galaxies (dSphs), satellites of the Milky Way, as the angular extent of these satellites is quite similar to nearby galaxies in the Local Group. This set of nearby galaxies and satellites has ideal conditions for the observation of DM annihilation and has been the subject of extensive theoretical and observational study \cite{mauroSearchGrayEmission2019, fengSearchingGeVGammaray2019, collaborationSearchingDarkMatter2015, collaborationObservationsMilkyWay2010}.

This paper is organized as follows. In sec.~\ref{sec:physics}, we summarize different scenarios in which $p$-wave and $d$-wave annihilation processes are dominant or comparable to $s$-wave channels. In sec.~\ref{sec:formalism} we detail our approach and methodology. In sec.~\ref{sec:results}, we present our findings. Lastly, we discuss the implications of our results in sec.~\ref{sec:discussion}.

\section{Velocity-Dependent Cross Sections}
\label{sec:physics}

We are interested in models in which the DM annihilation cross section has a non-trivial velocity dependence. In a model-independent approach, we parameterize this dependence as

\begin{equation}
    \label{eq:sigv}
    \langle \sigv \rangle = \langle \sigv \rangle_0 \Big(\frac{v_{\mathrm{rel}}}{c}\Big)^n.
\end{equation}
where $\sigma_A$ is the annihilation cross section and $\langle \sigv \rangle_0$ is the velocity-independent piece of the thermally averaged cross section. 

We aim to keep our approach as general as possible, but we will primarily consider three cases for illustrative purposes:

\begin{enumerate}
    \item $n = 0$: This is the $s$-wave channel, which is the typical case of a velocity-independent cross section;
    
    \item $n = 2$: The $p$-wave channel, which is relevant in particular models such as Majorana fermion DM annihilating into fermion/anti-fermion pairs. In this scenario, the $s$-wave channel is chirality-suppressed and so $p$-wave annihilation is dominant \citep{kumarMatrixElementAnalyses2013};
    
    \item $n = 4$: This is the $d$-wave channel, which becomes relevant in the case of real scalar singlet DM annihilating into lepton/anti-lepton pairs \citep{giacchinoScalarDarkMatter2013}. Again, the $s$-wave channel is chirality suppressed and now the $p$-wave channel requires a $CP$-odd bilinear involving two real scalars, for which there is no such operator. Similarly, for the case of scalar DM annihilation into a pair of massless gauge bosons, the cross section is also $d$-wave suppressed \citep{hanDiphotonResonanceGravity2016}.
\end{enumerate}

Sommerfeld enhancement ($n = -1$) is also a well-motivated case, but will tend to {\em enhance} the J-factor towards the center of a DM halo where the typical velocity is lower. We do not expect this case to yield a significantly different optimal observing strategy from the benchmark $s$-wave case. 
As such, we will exclusively focus on the 3 aforementioned cases.

\section{Formalism}
\label{sec:formalism}

\subsection{Velocity-Dependent J-Factor}
\label{subsec:velocityJfac}

The dark matter annihilation signal is proportional to the square of the DM density integrated over the line-of-sight. The literature refers to this integral as the J-factor, also called the astrophysical factor. In the velocity-independent case, it can be understood as a measure of how many DM pairs exist between an observer and their observation target. 
In the more general case, a velocity dependence can appear through the cross section, and the appropriate J-factor, which we indicate with the symbol $\jfac$, is defined as

\begin{equation}
    \jfac = \int dl \frac{\langle \sigv \rangle} {(\sigv)_0} \rho(r)^2,
\end{equation}
where $\rho$ is the DM density. Parameterizing the velocity-dependent annihilation cross section as in (\ref{eq:sigv}), the J-factor scales as the moments of the dark matter velocity \citep{boardVelocitydependentJfactorsAnnihilation2021}.

\begin{equation}
    \jfac = \int dl \rho(r)^2 \Big( \frac{\mu_n(r)}{c^n}\Big),
\end{equation}
where $\mu_n = \int d^3 \vrel f(\vrel) \vrel^n$ is the \textit{n}-th moment of the velocity distribution. 

We assume that the dark matter follows a Maxwell-Boltzmann distribution

\begin{equation}
    f(\mathbf{v}) \propto (\sigma_v^2)^{-3/2} e^{- \mathbf{v^2}/\sigma_v^2},
\end{equation}
where the velocity dispersion is $\sigma_v^2 = \frac{\langle v^2 \rangle}{3}$ from the equipartition theorem and we take $ \langle v^2 \rangle$ to be the square of the circular velocity at a given radius,

\begin{equation}
\label{eq:v}
v_c(r) = \sqrt{\frac{2 G M(<r)}{r}}.
\end{equation}

The average relative velocity of two dark matter particles of identical mass, assuming uncorrelated velocities,  is then given by 

\begin{multline}
    \langle \vrel^2 \rangle = \langle (\mathbf{v} - \mathbf{v}')^2 \rangle = \int d^3 \mathbf{v} \int d^3 \mathbf{v}' (\mathbf{v} - \mathbf{v}')^2 f(\mathbf{v}) f(\mathbf{v}') \\ = 2 \langle v^2 \rangle = 2 v_c^2
\end{multline}

For the purposes of this paper, we will consider two cases for the DM density: a ``cuspy'' Navarro-Frenk-White (NFW) profile and a ``cored'' Burkert profile  \citep{navarroUniversalDensityProfile1997a, bergstromDarkMatterGammaRays2006, burkertStructureDarkMatter1995}. Note that once we specify the density profile, the velocity is fully determined in this simple model, in which we neglect the gravitational potential of baryons. The NFW profile is given by \citep{navarroUniversalDensityProfile1997a}

\begin{equation}
    \rho (r) = \frac{\rho_0}{\left(\frac{r}{r_s}\right)\left(1 + \frac{r}{r_s}\right)^2}
\end{equation}
The mass enclosed in a sphere of radius $r$ centered on a halo with an NFW density profile is 

\begin{equation}
\label{eq:menc}
    M(<r) = 4 \pi \rho_0 r_s^3 \left( \frac{r_s}{r_s + r} - 1 + \log\left({1 + \frac{r}{r_s}}\right) \right).
\end{equation}
The calculation for the Burkert profile is analogous. Plugging (\ref{eq:menc}) into (\ref{eq:v}), we have everything we need to calculate the velocity-dependent J-factor.

\subsection{J-factor Calculation}

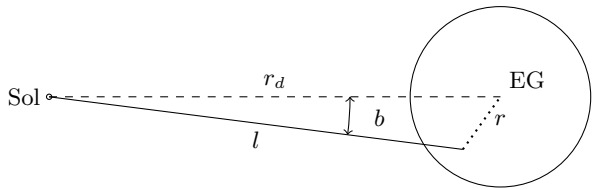
\begin{figure}
\centering
\begin{tikzpicture}

\coordinate (a) at (0,0);
\coordinate (b) at (6,0);
\coordinate (c) at (5.5,-0.7);

\draw (a) circle (1pt) node[left]{Sol};

\draw [dashed] (a) -- node[above] {$r_d$} ++(b);

\draw (a) -- node[below] {$l$} ++(c);

\pic["$b$", draw=black, <->, angle eccentricity=1.1, angle radius=4cm] {angle = c--a--b};

\draw (b) circle (1.2cm) node[above right]{EG};

\draw [dotted, thick] (c) -- (b) node[label=below:{$r$}] {};

\end{tikzpicture}
\caption{Geometric parameters used in J-factor calculation.  EG is a generic extragalactic body that hosts DM; $r_d$ is the displacement between our star, Sol, and the center of EG; $l$ is the displacment between Sol and a point in EG's DM halo; $b$ is the angle made between $r_d$ and $l$; and $r$ is the radial distance from the center of EG.}
\label{fig:geometry}
\end{figure}

Assuming a particular density profile, the DM velocity is known at all locations under the assumptions outlined in sec.~\ref{subsec:velocityJfac}. We can therefore perform the line-of-sight integral to calculate the J-factor per unit solid angle as a function of the angle $b$, which is measured with respect to the center of the target halo as shown in Fig.  \ref{fig:geometry}.

Expressing the distance to the center of a DM halo in terms of the distance from the solar neighborhood, we have 

\begin{equation}
    r = \sqrt{r_d^2 - 2l r_d \cos(b) + l^2},
\end{equation}
where $r_d$ is the distance between the Sun and the center of the DM halo and $l$ is the distance from the Sun and the point of evaluation. We then calculate the J-factor for a range of discrete values for $b$ from $0$ out to a maximum angle $b_{\mathrm{max}} = \arctan(\frac{r_t}{r_d})$, corresponding to a line tangent to the edge of the halo as defined by the tidal radius $r_t$. We interpolate between these points to get $\jfac(b)$. 

Because the ratio $r_s/r_d$ is nearly identical for M31, M87, and dSphs such as Draco, the angular scale of these halos is quite similar. We will therefore categorize all such halos  as an \textit{extragalactic} (EG) source case study. Note that we do not necessarily expect the shape of the DM distribution to be identical in each of these halos, but it is not necessary to separate these cases further for the illustrative purposes of this work. Thus for the EG case, we use model values corresponding to the dSph Draco: $\rho_0 = 2.3 \times 10^8 M_{\odot} \mathrm{kpc}^{-3}$,  $r_d = 76$ kpc, $r_t = 0.97$ kpc, and $r_s = 0.35$ kpc as calculated in \citep{klopImpactAxisymmetricMass2017}. For the Galactic Center, we use $\rho_0 = 4.9 \times 10^6 M_{\odot} \mathrm{kpc}^{-3}$, $r_d = 8$ kpc, $r_t = 90$ kpc, and  $r_{s} = 15.3$ kpc. 

\begin{figure}
\centering
\begin{tikzpicture}

\coordinate (a) at (0,0);
\coordinate (b) at (1.5,0);
\coordinate (c) at (0,1);

\draw[pattern=north west lines, pattern color=blue] (a) circle (1.5);

\draw[fill=white] (a) circle (1);

\draw (a) -- node[below] {$\theta_2$} ++(b);

\draw (a) -- node[left] {$\theta_1$} ++(c);

\end{tikzpicture}
\caption{The shaded annulus region is the area in which we evaluate the signal-to-noise ratio.}
\label{fig:integralgeometry}
\end{figure}
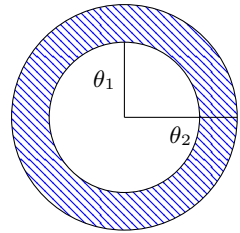

To determine the total J-factor for a particular solid angle field of view, we integrate the J-factor over all relevant angles, as shown in Fig. \ref{fig:integralgeometry}. For an annulus centered at $b=0$, making use of the spherical symmetry of the halo, the total J-factor is

\begin{equation}
    2 \pi \int_{\theta_1}^{\theta_2} \mathcal{J}(b) b d b.
\end{equation}

\subsection{Gamma-Ray Background}

One of the challenges in conclusively detecting a signal from dark matter annihilation is that the signal must be distinguishable from the astrophysical background. At gamma-ray energies,  key sources of background include inelastic  cosmic-ray interactions with ionized gas and inverse Compton scattering of high-energy electrons off of the intervening photon background. Gamma-ray point sources within or beyond the halo of interest contribute additional noise.

There are certain characteristics of DM annihilation that may make it distinguishable from the astrophysical background. In addition to model-dependent spectral features, DM annihilation is expected to result in a signal that is highly concentrated towards the central regions of the halo, but still more extended than a point source. In order to extract a potential signal, it is paramount to identify the optimal observation region, which, we define here as the region with the largest signal-to-noise ratio.

Regarding gamma-ray sources within the DM halos of interest, studies using data from the Fermi Large Area Telescope (LAT) have shown that no statistically significant excess of gamma-ray emission is present in 8 observed dSph candidates \cite{drlica-wagnerSEARCHGAMMARAYEMISSION2015}. However, in larger galaxies, a significant gamma-ray background from the central regions of the halo has been detected. See, for instance, promising targets for DM searches such as the M31 (Andromeda) galaxy, where a central diffuse emission has been detected \cite{karwinDarkMatterInterpretation2021, doCosmicrayTransportGammaray2020}; in other cases, the background source is genuinely a point-like object, such as an active galactic nucleus as in the case of M87 \cite{thefermilatcollaborationFermiLargeArea2009}.

For simplicity, we assume a model of  gamma-ray background consisting of two components. For the extragalactic case study, we use (1) an isotropic component, modeling the diffuse background, and (2) a central point source, suitably smeared to account for the finite instrumental point spread function (see details below). The isotropic piece is aimed at capturing both the diffuse gamma-ray background produced by interactions of high energy cosmic rays with interstellar gas, and the extragalactic gamma-ray background. The point source accounts for the contribution from gamma-ray emission in the central region of galaxies from active galactic nuclei or other sources, such as a dense population of unresolved millisecond pulsars \cite{bartelsStrongSupportMillisecond2016, hooperMillisecondPulsarsTeV2018}. 

For the Galactic Center (GC), we expect that both the DM annihilation and other astrophysical processes would result in spherically symmetric emission. Data from the Fermi Gamma Ray Space Telescope show that gamma-ray emission close to the Galactic Center is distributed as $r^{-1.55}$, where $r$ is the distance to the Galactic Center \citep{hooperDarkMatterAnnihilation2011}. This is perhaps due to the gas or inverse Compton emission in the Galactic Bulge or from a population of unresolved point sources.  We expect the isotropic background to be subdominant for the Galactic Center, and therefore ignore it. In this case then, the two components of the background are (i) $N_{B} \propto \int dl r^{-1.55}$, and again (ii) a central point source, possibly associated with the central supermassive black hole Sag A$^*$ or a concentrated cluster of faint point sources.

We model the contribution from the point source as a Gaussian centered at the center of the given DM halo ($b = 0$). The photon count from this component of the background is proportional to 

\begin{equation}
    N_{G}(b) \propto \frac{1}{\sigma \sqrt{2 \pi}} e^{-\frac{1}{2} \big(\frac{b^2}{\sigma^2} \big)},
\end{equation}
such that $\sigma$ is the angular width of the central source accounting for the instrument-dependent point spread function. We define $N_{G}(b)$ as the contribution from the entire line-of-sight for a given angle $b$ so that the total background from the point source is integrated over the field of view

\begin{equation}
    N_{Gtot} \propto \int_{\theta_1}^{\theta_2} N_{G}(b) b db.
\end{equation}

The total isotropic contribution to the background is simply proportional to the field of view

\begin{equation}
    N_{Itot} \propto N_{I0}(\theta_2^2 - \theta_1^2),
\end{equation}
where $N_{I0}$ is the isotropic background per solid angle. Thus the total background contribution is simply $N_{Itot} + N_{Gtot}$.

We normalize the isotropic background relative to the area of the Gaussian point source (i.e. 1) so that $N_{G tot} = \eta N_{I0}$, in which $\eta$ can take on a range of values. For the Galactic Center case, we normalize the bulge component of the background relative to the point source, such that $N_{G tot} = \eta N_{B}(0)$. For the Galactic Center, the total background is thus $N_{B tot} + N_{G tot}$ where $N_{B tot} \propto \int_{\theta_1}^{\theta_2} N_{B}(b) b db$.

A large $\eta$ corresponds to the case where the central gamma ray source dominates the background, such as in the case of galaxies with an active galactic nucleus. Conversely, a smaller $\eta$ corresponds to a smaller contribution from the central region of a halo and a flatter background profile. This is most relevant for dSphs where there is not typically a large central gamma ray source, but a sub-dominant faint source below detection threshold cannot be ruled out (see e.g. \cite{gonzalez-moralesEffectBlackHoles2014a}).

The number of detected photons amounts to counting independent events that randomly occur at a constant rate, and is therefore well described by a Poisson distribution. Since the mean value of a Poisson distribution is also its variance, the standard deviation is simply the square root of the count. We therefore take the noise to be given by the square root of the number of background gamma-ray events. Therefore, we define the ``{\em optimal observing region}'' as the region that maximizes the quantity

\begin{equation}
    \frac{\jfac}{N_\mathrm{{tot}}^{1/2}}.
\end{equation}

\section{Results}
\label{sec:results}

\subsection{Extragalactic Source}

\begin{figure}
    \centering
    \includegraphics[scale=0.5]{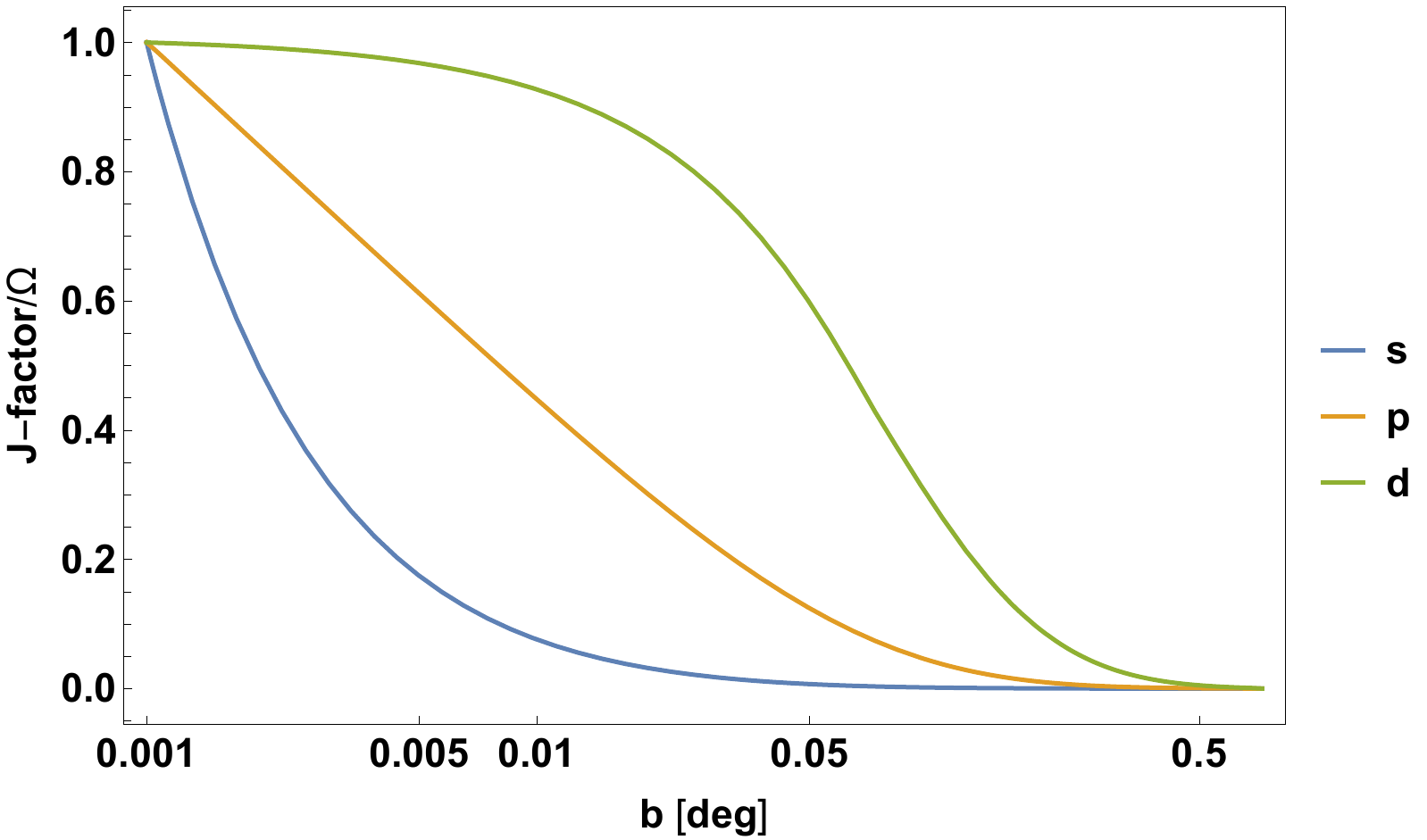}
    \caption{$\jfac$ per solid angle as a function of the angle $b$ for different annihilation channels, normalized to their respective values at $b=10^{-3}$ degrees. The contribution to the J-factor from the velocity-dependent $p$ and $d$-wave channels extends to larger angles when compared to the velocity-independent $s$-wave case.}
    \label{fig:losAll}
\end{figure}

For the extragalactic case, we show $\jfac(b)$ per unit solid angle for $s$-wave, $p$-wave, and $d$-wave annihilation in Fig.\ref{fig:losAll}. Note that due to the assumption of the halo's spherical symmetry, $\jfac$ is only a function of one variable, the angle $b$. 

In the $s$-wave case, the J-factor per unit solid angle decreases rapidly with increasing $b$, decreasing approximately to $90 \%$ from $b = 0.001$ to $b = 0.01$ degrees. By comparison, the $p$- and $d$-wave channels decrease at more modest rates and contribute significantly to the J-factor out to larger angles. This is because velocity increases as a function of $r$, the distance from the center of the halo, up until $r \approx r_t$ at which point the DM particles are tidally stripped away and the enclosed mass ceases to increase. Thus, the line-of-sight integral through the halo will be enhanced when including this higher velocity region due to the velocity-dependence of the cross section. Because the noise grows as the square root of the background, there must be a significant enhancement of $\jfac$ or a large spike in the background signal at small $b$ in order for it to be prudent to exclude the central region from the region of interest. We note that there are uncertainties associated with the calculation of the tidal radius. However, only a small fraction of the total annihilation flux comes from the outermost regions of the halo, so we don't expect changes in the tidal radius to affect our results in any meaningful way (see the discussion in \cite{klopImpactAxisymmetricMass2017} for more details).

\begin{figure*}
    \centering
    \includegraphics[scale=1.0]{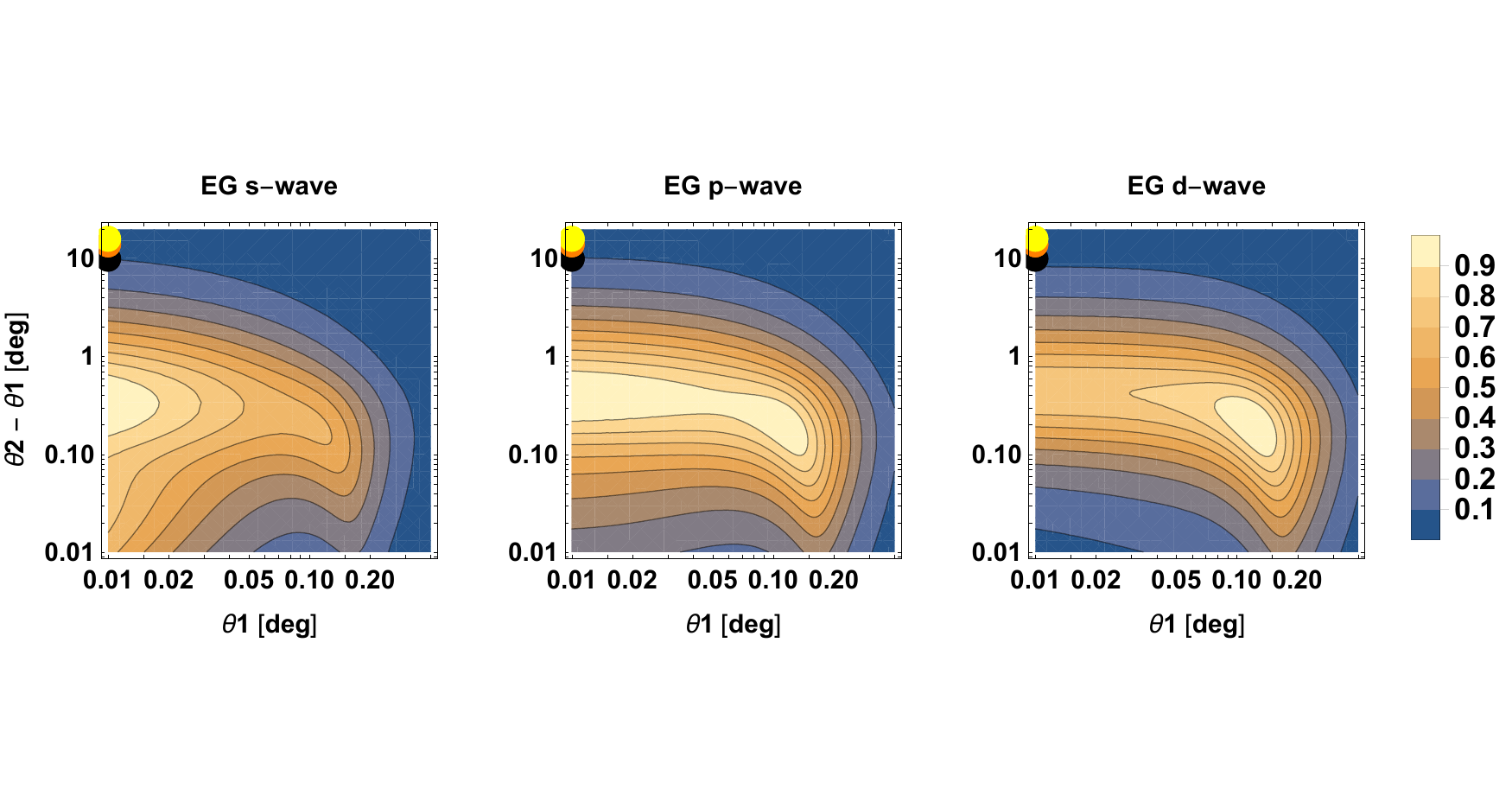}
    \vspace{-0.8in}
    \caption{$ \frac{\jfac}{N_\mathrm{{tot}}^{1/2}}$ for $s$, $p$, and $d$-wave annihilation for an extragalactic source using an NFW density profile with $\sigma = 0.05^\circ$ and $\eta = 20$. The black dot is the field of view used in a Fermi-LAT survey of MW satellites \cite{fermi-latSearchingDarkMatter2017, collaborationObservationsMilkyWay2010}, while the orange and yellow dots correspond to observations of M31 using Fermi-LAT \cite{fengSearchingGeVGammaray2019, mauroSearchGrayEmission2019}.}
    \label{fig:EGNFWFinalFig}
\end{figure*}

We plot $\frac{\jfac}{N_\mathrm{b}^{1/2}}$ as a function of $\theta_1$ and $\theta_2 - \theta_1$, the angular extent of the inner side of the annulus and the thickness of the annulus, respectively. Using an NFW density profile for Draco, we show the signal-to-noise ratio for a dominant central gamma-ray source ($\eta = 20$) for the case of $s$-wave, $p$-wave, and $d$-wave annihilation in Fig. \ref{fig:EGNFWFinalFig}. We see that for the $s$-wave channel, the optimal strategy is a large circular field of view, including the entire angular extent of the source. For the $p$-wave case, there is a {\em much broader} range of observation strategies that yield large signal-to-noise ratios; the disk field of view is still viable, but excluding the central region yields a similar outcome. However, for the $d$-wave case, the optimal observation strategy becomes an annulus, excluding the inner $0.1$ degrees of the halo and having a slightly smaller field of view, with a width extending only about $0.3$ degrees, i.e. cropping out both the inner and outer regions of the halo. 

For reference, the dots shown on the EG figures represent the observation strategies used in previous searches for gamma-ray emission due to dark matter annihilation. The black dot corresponds to surveys of Milky Way dSphs using Fermi-LAT data \citep{collaborationObservationsMilkyWay2010, fermi-latSearchingDarkMatter2017}. The orange and yellow dots correspond to observations of M31, again using Fermi-LAT \cite{mauroSearchGrayEmission2019, fengSearchingGeVGammaray2019}. We note that a square field of view (e.g. $10^\circ \times 10^\circ$) is represented as a circular field of view (e.g. $\theta_2 - \theta_1 = 10^\circ$) in the figure for ease of comparison. 

It is also worth keeping in mind that the angular resolution of Fermi-Lat varies depending on the single-photon energy. For $>10$ GeV photons, the resolution is $<0.15^\circ$, whereas for sub-GeV photons, the resolution is $\gtrsim 1.0^\circ$. The typical energy range observed is $\approx 100$ MeV - $500$ GeV. For comparison, next generation telescopes, such as AdEPT, should achieve $0.1^\circ$ at $\approx 1$ GeV \cite{hunterPairProductionTelescope2014}. We see that even with current-generation instruments, there is room for improvement in terms of the signal-to-noise ratio for the given astrophysical background.

\begin{figure*}
    \centering
    \includegraphics[scale=0.7]{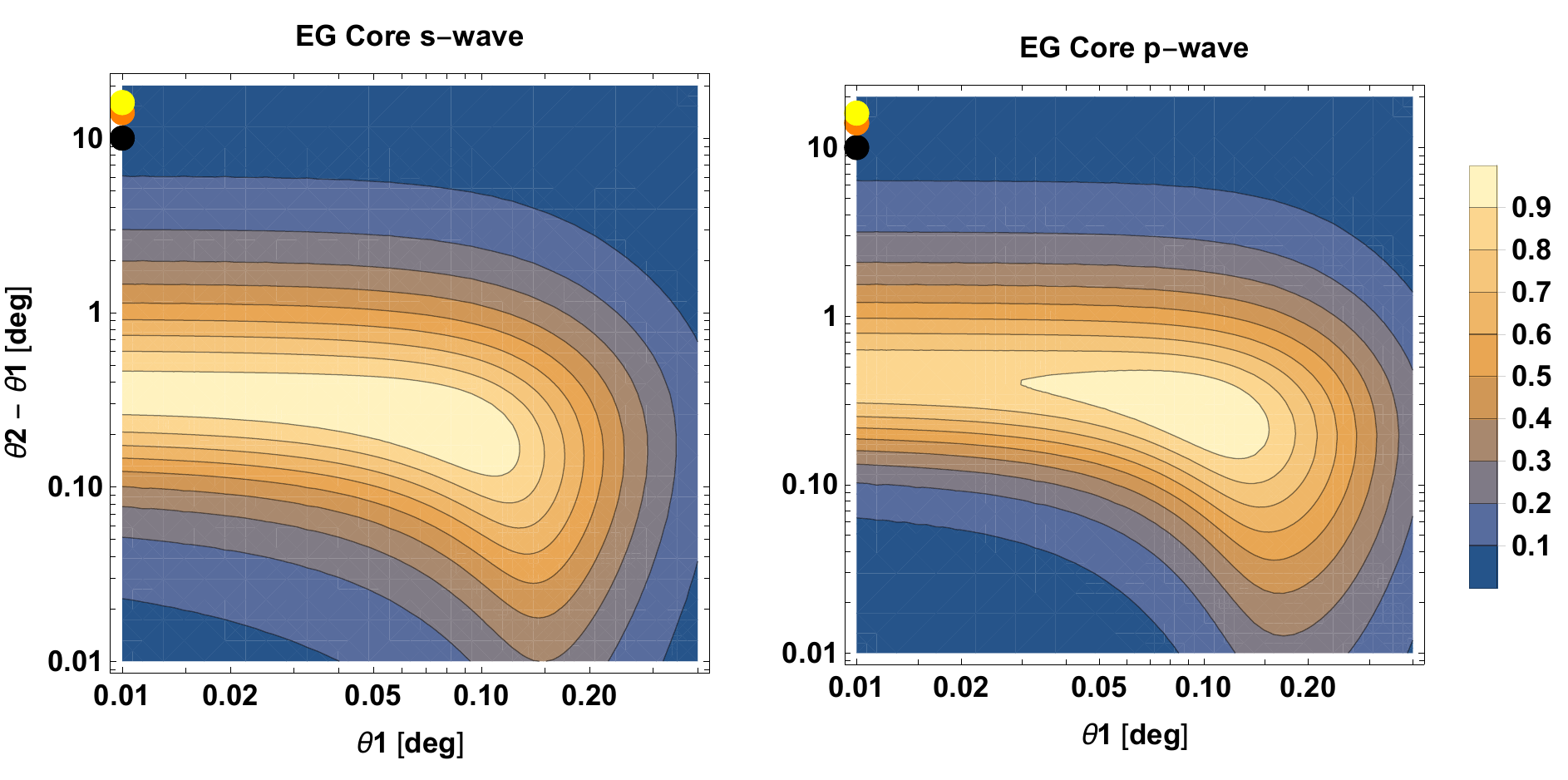}
    \caption{$ \frac{\jfac}{N_\mathrm{{tot}}^{1/2}}$ for $s$ and $p$-wave annihilation for an extragalactic source using a Burkert profile with $\sigma = 0.05^\circ$  and $\eta = 3$. The black dot is the field of view used in a Fermi-LAT survey of MW satellites \cite{fermi-latSearchingDarkMatter2017, collaborationObservationsMilkyWay2010}, while the orange and yellow dots correspond to observations of M31 using Fermi-LAT \cite{fengSearchingGeVGammaray2019, mauroSearchGrayEmission2019}.}
    \label{fig:EGCoreFinalFig}
\end{figure*}

Using the cored Burkert density profile, 

\begin{equation}
    \rho_{\mathrm{c}} = \frac{\rho_0}{\big(1 + \frac{r}{r_s}\big) \big(1 + (\frac{r}{r_s})^2 \big)}
\end{equation}
we show the signal to noise ratio in Fig. \ref{fig:EGCoreFinalFig} for $s$- and $p$-wave channels. The behavior of the $d$-wave channel in this case is quite similar to that of the $p$-wave channel so we only show $s$ and $p$ for simplicity. The width of the central point source is fixed at $\sigma = 0.05$ degrees. Similarly to the corresponding NFW profile examples, the best field of view for the velocity-independent channel is either a disk or a thin annulus. For the velocity-dependent channels, the optimal field of view is achieved with an annulus, excluding the inner $0.05$ degrees of the halo, which corresponds to the width of the point source. Note that the point source need not be extremely bright for this to be the case; this behavior occurs for a moderate value of $\eta = 3$. The extent of the annulus is only about $0.4$ degrees, whereas the total extent of sources like M31, M87, and Draco are about $0.7$ degrees, so excluding the outermost region of these halos would be beneficial. 

\subsection{Galactic Center}

\begin{figure*}
    \centering
    \includegraphics[scale=0.7]{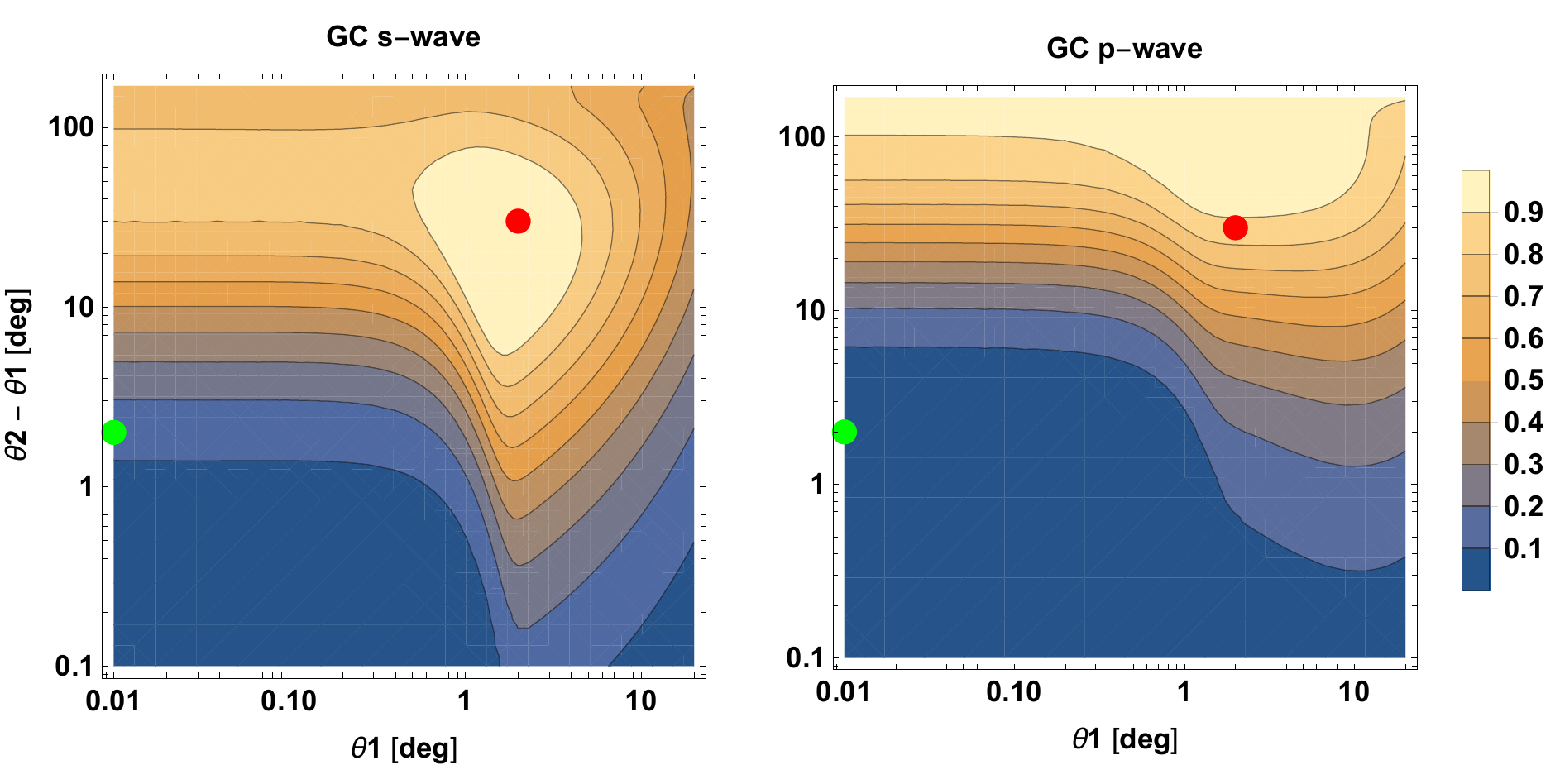}
    \caption{$\frac{\jfac}{N_\mathrm{{tot}}^{1/2}}$ for $s$ and $p$-wave annihilation in the Galactic Center with $\sigma = 0.5^\circ$ and $\eta = 50$, corresponding to a dominant central point source. The green dot is the field of view used by a Fermi-LAT survey of the Galactic Center looking for $p$-wave annihilation \cite{johnsonSearchGammarayEmission2019}. The red dot represents a previous search for $s$-wave annihilation using Fermi-LAT with a 2 degree plane mask \cite{leaneDarkMatterStrikes2019}.}
    \label{fig:GCRatio50FinalFig}
\end{figure*}

\begin{figure*}
    \centering
    \includegraphics[scale=0.7]{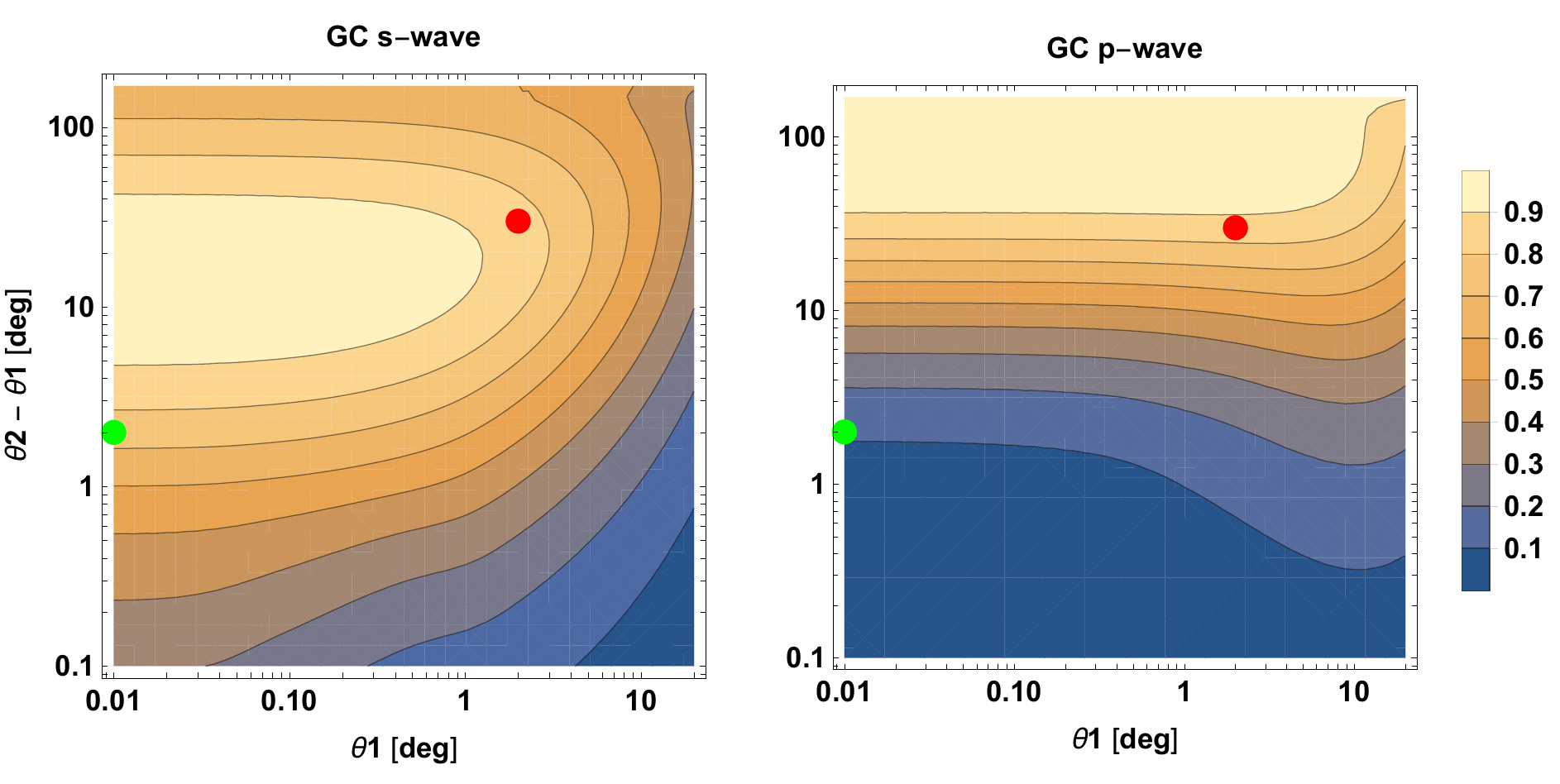}
    \caption{$\frac{\jfac}{N_\mathrm{{tot}}^{1/2}}$ for $s$ and $p$-wave annihilation in the Galactic Center with $\sigma = 0.5^\circ$ and $\eta = 0.1$, corresponding to a dominant noise contribution from the GC bulge. The green dot is the field of view used by a Fermi-LAT survey of the Galactic Center looking for $p$-wave annihilation \cite{johnsonSearchGammarayEmission2019}. The red dot corresponds to a search for $s$-wave annihilation using Fermi-LAT with a 2 degree plane mask \cite{leaneDarkMatterStrikes2019}.}
    \label{fig:GCRatio01FinalFig}
\end{figure*}

We calculate $\frac{\jfac}{N_\mathrm{b}^{1/2}}$ for the Galactic Center for $s$- and $p$-wave annihilation under two scenarios: In the first, the central gamma-ray source is large compared to the bulge component with $\eta = 50$, a case shown in Fig. \ref{fig:GCRatio50FinalFig}. In the second scenario, the central gamma-ray source is very small compared to the bulge component with $\eta = 0.1$, a case shown in Fig. \ref{fig:GCRatio01FinalFig}. The second case can be thought of as corresponding to an optimistic scenario where the bulk of the central Galactic emission stems from dark matter, versus a pessimistic one where it does not (originating instead from e.g. unresolved point sources).

Let us first consider the case with a bright central point source, shown in Fig. \ref{fig:GCRatio50FinalFig}. For the $s$-wave channel, the highest signal-to-noise ratio can be obtained by using an annulus, blocking out the innermost degree and extending out to tens of degrees. Indeed, this is similar to the approach taken by \cite{leaneDarkMatterStrikes2019} which uses a plane mask for the inner 2 degrees of the Galactic latitude and a field of view extending out to 30 degrees. This approach is represented as a red dot in Figs. \ref{fig:GCRatio01FinalFig} and \ref{fig:GCRatio50FinalFig}. Note that the present work is done considering a purely spherical halo, so the inner ring of the annulus would not be confined to the plane of galactic latitude. Even still, it is interesting to find that an existing analysis uses a mask ideal for detecting $s$-wave annihilation according to this calculation.

The $p$-wave case yields a very different result. Here, the best approach is to include as wide a region as possible. This is because the velocity enhancement of the cross section leads to a non-negligible contribution to the J-factor per solid angle, even far from the Galactic Center. In fact, one can exclude as much as the inner 10 degrees of the Galactic Center, and still end up with an ideal signal to noise ratio by using a very wide field of view. 

The dots shown in the GC figures represent the observation strategies used in previous gamma-ray surveys in the Galactic Center. The red dot represents a survey using Fermi-LAT data for energies $2-20$ GeV with a $< 2^\circ$ plane mask with respect to the Galactic latitude \cite{leaneDarkMatterStrikes2019}. For ease of comparison, we represent this as a circular $2^\circ$ mask in the figures. Interestingly. we find that the plane mask yields a very high signal-to-noise ratio for both the $s$- and $p$-wave scenarios given the astrophysical background used here. This suggests that one does not necessarily need to search for a particular annihilation channel at the exclusion of another. 

The green dot corresponds to a search for $p$-wave annihilation of dark matter using Fermi-LAT in the energy range $10-600$ GeV \cite{johnsonSearchGammarayEmission2019}. The region of interest was $2^\circ \times 2^\circ$, approximately represented in the figure as $\theta_2 - \theta_1 = 2^\circ$. Here, the goal was to make use of the increase in the DM velocity in the region of the central supermassive black hole Sgr $A^*$. For this region of interest, we find a relatively low signal-to-noise ratio for both $s$ and $p$-wave annihilation. However, we note that the corresponding survey considered the existence of a localized dark matter over-density near Sag $A^*$, which could significantly change the J-factor \cite{johnsonSearchGammarayEmission2019}. We did not consider DM substructure in this work, so we leave this possibility to future investigation. 

We now turn to the second GC case, shown in Fig. \ref{fig:GCRatio01FinalFig}, which has a small central gamma-ray source compared to the bulge component. Here, the optimal strategy for the $s$-wave channel is once again a disk. With no dominant point source, the highest signal-to-noise ratio is attained by including the center of the halo where the dark matter density is highest. Again in the $p$-wave case, the velocity enhancement implies that a much larger field of view should be used, regardless of whether or not the inner region of the halo is included in the field of view.

\section{Discussion and conclusions}
\label{sec:discussion}

In this paper, we have investigated the behavior of the DM annihilation rate through its dependence on the J-factor for cases in which the annihilation cross section is velocity-dependent. Under simple assumptions for the astrophysical background, we find that the annihilation signal-to-noise ratio can be larger for an annulus field of view than for a disk when considering $p$- and $d$-wave channels, even when there is not an overwhelming background from a central gamma-ray point source. This can be true for the extragalactic case, regardless of whether the DM distribution is cuspy or cored. In the case of the Galactic Center, the optimal observation strategy for velocity-dependent channels is again quite different than for the velocity-independent one, but this manifests differently than in the extragalactic scenario. Here, we find that the $p$-wave signal-to-noise ratio is optimized for a much larger field of view than that of the $s$-wave signal. Intriguingly, one can search for both channels simultaneously and still have a very high signal-to-noise ratio for each by masking the inner 1-5 degrees of the halo and extending the field of view out to nearly 100 degrees.

We have taken a number of simplifying assumptions throughout this paper for illustrative purposes. It is now worthwhile to mention and revisit these assumptions. We have neglected the baryonic matter contribution to the gravitational potential, which would contribute to the DM's velocity dispersion (this is expected to be a subdominant effect in some cases, for instance dSph, but more relevant in others, such as the Galactic Center). We  have also ignored the possibility of DM substructure within the Milky Way and extragalactic halos. Such substructures would alter both the density and the  velocity distribution of the dark matter, and therefore that would go beyond our analysis. 

Another simplifying assumption was that the DM relative velocity distribution follows a Maxwell-Boltzmann distribution. In reality, this is an oversimplification as the distribution is only truly Maxwell-Boltzmann for the case of a singular isothermal sphere. More generally, one could consider an anisotropic distribution and relate the DM velocity distribution to the density profile through the Eddington inversion formula \citep{widrowDistributionFunctionsCuspy2000}. In this case, the velocity dispersion would no longer be trivially related to the circular velocity and one would need to more accurately determine the relative velocity of the dark matter particles to calculate $\jfac$. 

Specific values for the width of the central gamma-ray point source $\sigma$ and the relative normalization of the background sources $\eta$ were chosen {\em ad hoc} on the basis that these served as reasonable benchmarks and demonstrated non-trivial results for various annihilation channels. The purpose of this more general work was to demonstrate how one might carry out such analysis and to demonstrate a few interesting cases as examples. Ideally, one would analyze a given target of observation and determine the optimal strategy for that particular target using the approach and method presented here. 

Lastly, we have examined both an NFW and core-type Burkert density profile for the dark matter. In reality, the density distribution of DM halos is wide ranging \citep[for the case of dSph see e.g.][]{omanUnexpectedDiversityDwarf2015}. As a result, the optimal observational strategy for different dSphs may vary, even for the same annihilation channel and similar central gamma-ray sources.

\begin{acknowledgments}
This material is based upon work supported in part by the National Science Foundation Graduate Research Fellowship under Grant No. DGE-1842400 to NS. SP is partly supported by the U.S.\ Department of Energy grant number de-sc0010107. 
\end{acknowledgments}

\section*{Data availability}
The code used in this article will be shared upon  request to the authors.


\bibliography{main}

\end{document}